\documentclass{pasj00}
\input psfig.sty

\def\etal{{\rm et~al.\ }} 
\def\HI{H\,{\sc i}\ }

\SetRunningHead{J. L. Han et al.}{The Fundamental Plane of Spiral Galaxies}

\title{The Fundamental Plane of Spiral Galaxies:
 \\Search from Observational Data}

\author{JinLin \textsc{Han},\altaffilmark{1,5,6}
        Zugan \textsc{Deng},\altaffilmark{2,5} 
	Zhenlong \textsc{Zou},\altaffilmark{1,5}
	Xue-Bing \textsc{Wu},\altaffilmark{3,5} and
        Yipeng \textsc{Jing}\altaffilmark{4} 
	} 

\altaffiltext{1}{National Astronomical Observatories, Chinese Academy 
of Sciences, Beijing 100012, China}
\email{hjl@bao.ac.cn, zzl@bao.ac.cn}
\altaffiltext{2}{Department of Physics, Graduate School, Chinese Academy 
of Sciences, Beijing 100039, China}
\email{dzg@bac.pku.edu.cn}
\altaffiltext{3}{Department of Astronomy, Peking University, Beijing 100871, China}
\email{wuxb@vega.bac.pku.edu.cn}
\altaffiltext{4}{Shanghai Astronomical Observatory, the Partner Group  
of MPI f\"ur Astrophysik,
 Shanghai 200030, China}  
\email{ypjing@center.shao.ac.cn}
\altaffiltext{5}{Beijing Astrophysical Center, CAS-PKU, Beijing 100871, China} 
\altaffiltext{6}{The partner group of
Max-Planck-Institute f\"ur Radioastronomy at National Astronomical
Observatories of China}


\Received{2001/06/28}
\Accepted{2001/07/31}
\Published{2001/10/25}                                  
\begin{document}
  
\maketitle

\begin{abstract} 
The fundamental plane of spiral galaxies was searched from observational
data, which can be approximately represented in observational parameters
by  $ L \propto V^2 R $, where $L$ , $R$, and $V$ are the luminosity,
the linear size of galactic disk, and the rotation velocity. This plane
exists at all optical bands in our samples of more than 500 spiral galaxies
in total. It is more fundamental than the relationship between any
two of the parameters, and can reduce the residual of the Tully--Fisher
$L$--$V$ relations by about 50\%. Noticed that the power index of $V$ is
doubled from that of $R$, which implies that the total mass and the
mass distribution of galaxies plays an important role in forming the
fundamental plane. Involving the third parameter of galactic size
has a strong physical implication on galaxy formation and the dark--mass
distribution.
\end{abstract} 

\section{Introduction} 
The Tully--Fisher relation is a tight correlation between the 
internal motion and the luminosity of spiral galaxies \citep{tf77}. It 
can be expressed as  
$$M = \alpha \log V + \gamma, \eqno(1)$$ 
where  $M$ is the absolute magnitude, $V$ the rotation velocity,
$\alpha$ the slope of the relation, and $\gamma$ the zero-point. 
This empirical relation has been used for estimating the distances of 
galaxies, and hence for determining the Hubble constant (e.g.
\cite{smh+00}; \cite{tp00}).  It also provides a critical 
constraint on galaxy formation (e.g. \cite{dss97};
\cite{mmw98}; \cite{vdb00}; \cite{af00}; \cite{fa00}; Koda et al. 
2000b\nocite{ksw00b}). However, the detailed physical processes
concerning the origin of the Tully--Fisher relation have not yet
been fully understood. This relation 
could be a result of the self--regulated star formation in the disks of 
different mass (see  \cite{sil97}; \cite{el96}), or a direct 
consequence of the cosmological evolution between the mass and the circular 
velocity (see e.g. \cite{sn99}; \cite{mmw98}; \cite{mm00}). 
 
It was believed that the residual value deviating
from the Tully--Fisher relation can pose strong constraints on scenarios
of galaxy formation and evolution (see e.g. \cite{el96}) which was
systematically studied by, for example, Willick et al. (\yearcite{wcf+97}).
The residuals, written as
$$\delta = M - (\alpha \log V +\gamma), \eqno(2)$$
are not accountable by measurement errors. With good data at the $R$ and $I$
bands, the scatter of the luminosity may be partially attributed to the
measurement uncertainty of 0.08 mag (\cite{cou96}) or 0.04 mag
(\cite{ghh+97}), which is further convolved with
worse-determined distances. Because of the residuals,
the estimates of the galaxy distance based on the Tully--Fisher relation 
have a typical uncertainty of 20\%.

Some efforts have been made previously to search for tighter correlations
among the luminosity, rotation velocity, and disk radius for spiral galaxies.
\citet{kod89} found a much tighter correlation among
these three parameters, $L  \propto V R^2$.  He pointed out that
the above relation is valid for ellipticals as well.
Very recently, Koda et al. (2000a)\nocite{ksw00a} found that the $I$-band 
luminosity $L$, $I$-band radius $R$, and the rotation velocity $V$ of
spiral galaxies are distributed on the so-called scaling plane in 
three-dimensional parameter's space of these quantities, which can be
expressed as $L \propto 
(VR)^{1.3}$. Because the galactic radius was considered (Koda et al.
2000a)\nocite{ksw00a}, the residual of the Tully--Fisher relation was
significantly reduced. In fact, \citet{cr99}
also realized the importance of involving the disk size to the Tully--Fisher
relation, but they did not come up with such a relation. \citet{wil99}
explored the role of the disk scale length, and found the surface brightness
dependence of the Tully--Fisher relation.

It is intriguing to know whether and how the third parameter, the galactic
size, really plays some role in galaxy formation, and how the mass and
luminosity are physically related to this parameter. Besides strong
evidence found by Koda et al. (2000a)\nocite{ksw00a},
the linear (optical) galactic diameter has a very tight correlation
to the H~{\sc i} mass in later galaxies \citep{br97}. Koda et al.
(2000b)\nocite{ksw00b} have already done a set of simulations. They
found that the galactic size was involved as a key parameter in such a
process that the
galactic mass and angular momentum were controlled during galaxy formation.
Very recently, \citet{sms01} considered the variations of the model parameters
presented in the theoretical work of Mo et al. (1998)\nocite{mmw98}. They
also found a theoretical fundamental plane, $L \propto V^{2.6} R^{0.5}$.

It is therefore very necessary to search the fundamental plane from
observational data with the galactic radius, $R$, as the third parameter,
and see how it matches the results from simulations of galaxy formation.
The primary procedure we take here is to see how
the galactic size can help to reduce the residual of the Tully--Fisher
relation. The best-fitting plane, expressed as
$$ M = \alpha \log V + \beta \log R  + \gamma, \eqno(3) $$ 
was searched from observational data with three variables,
$\alpha$, $\beta$, and $\gamma$.  Koda et al. (2000a)\nocite{ksw00a}
obtained  $\alpha=\beta=-3.25$, while Kodaira (1989)\nocite{kod89}
gave $\alpha=-2.5$ and $\beta=-5.0$. We found  that the best-fitting
plane with  $\beta/\alpha$ around 0.5 for different data-sets, namely,
the $I$-band data of Han (1992)\nocite{han92} and Palunas and Williams 
(2000)\nocite{pw00}, the $R$-band data of Courteau (1996, 
1997)\nocite{cou96}\nocite{cou97}, and the $BVRIH$-band data of calibration 
galaxies in Sakai et al. (2000) \nocite{smh+00} and Macri et 
al. (2000)\nocite{mhs+00}. The plane exists in different
wavebands, and can reduce the residual of the Tully--Fisher relation
by about 50\%. This implies that about 50\% of the scattering
in the observed Tully--Fisher relations is dominated by the intrinsic
properties of galaxies, and the other 50\% by measurement
uncertainties. We believe that the  plane is fundamental for spiral
galaxies, similar to that for elliptical galaxies 
(\cite{dd87}; \cite{dlb+87}). It should provide useful 
constraints on models of galaxy formation (see \cite{sms01}; Koda et al. 
2000b\nocite{ksw00b}).

\begin{figure*}[t]		
   \begin{center}
\centerline{\psfig{file=JLHan.fig1a.ps,width=175mm,height=70mm,angle=270}} 
\centerline{\begin{tabular}{cc} 
\mbox{\psfig{file=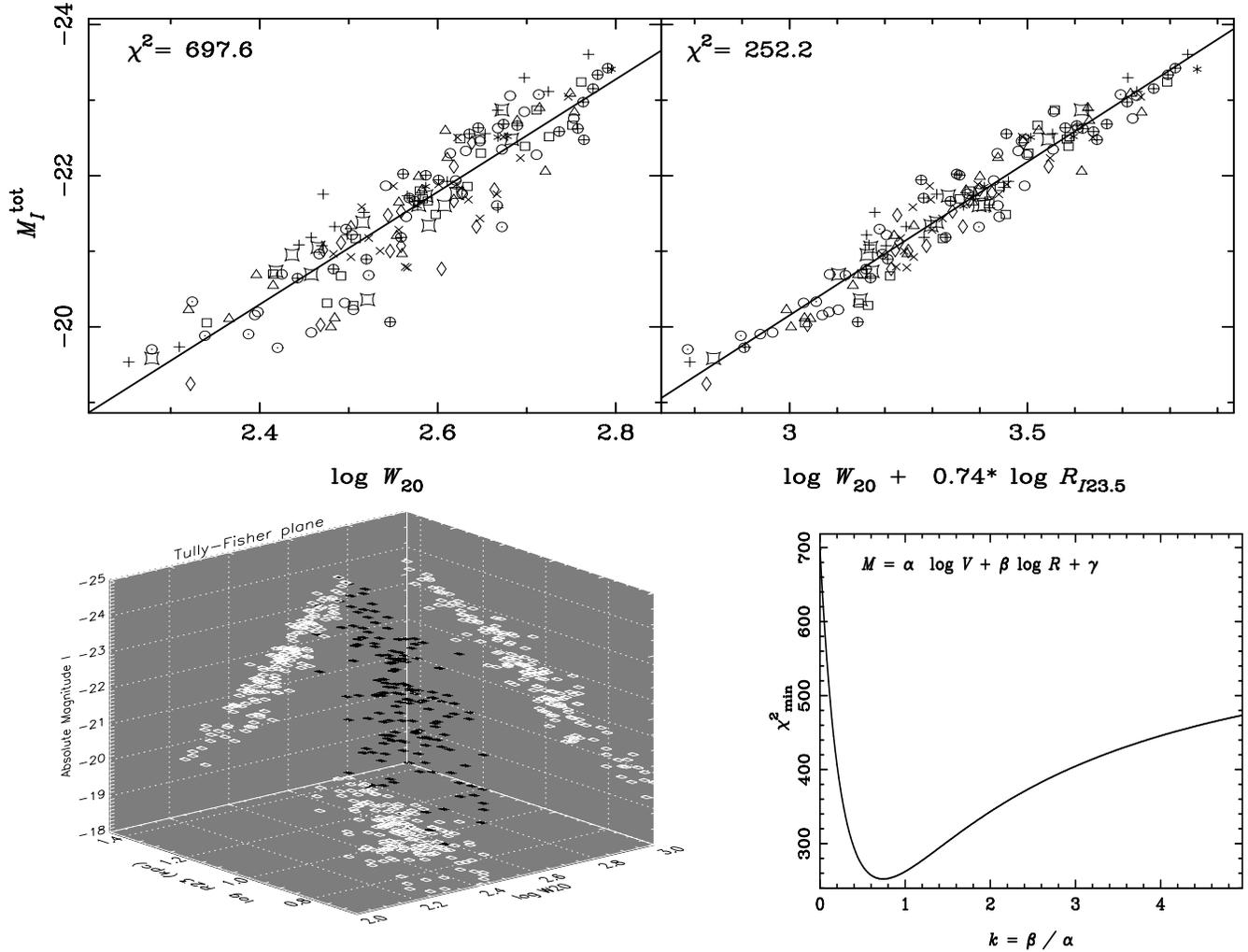,width=105mm,height=65mm}} & 
\mbox{\psfig{file=JLHan.fig1c.ps,width=68mm,height=60mm,angle=270}}
\end{tabular}} 
   \end{center}
\caption{Comparison of the Tully--Fisher relation (the left side) and 
the best fitting plane (the right side). Data of 160 galaxies were taken from 
Han (1992). The typical rms errors of the absolute magnitudes, together 
with the displacement 
from the uncertainties of $W_{\rm 20}$ and $R_{\rm 23.5}$, were taken 
as 0.2 mag. Galaxies in different clusters were plotted by different 
symbols. The data distribution in 3-D space and the variation of fitting
$\chi^2$ with $k$ were also plotted.
\label{han92chi} 
} 
\end{figure*} 

\section{The Fundamental Plane versus the Scaling Plane} 
 
Han (1992)\nocite{han92} presented carefully-calibrated $I$-band data 
for galaxies in clusters, including the total magnitude, $M_I^{\rm tot}$ and 
$M_I^{23.5}$ (in units of mag), and the face-on $I$-band isophotal radius,
$A^0_{23.5}$ (in units of arcsec). He also gave the width of the 21 cm 
\HI\ line, $W_{20}$ (in units of km~s$^{-1}$) corrected for the 
redshift and inclination.  The scaling plane, $L \propto 
(VR)^{1.3}$ (see figure~1 of Koda et al. 2000a\nocite{ksw00a}), was
found among these three  observational parameters. Therefore, the
Tully--Fisher relation between $M_I$ and 
$W_{20}$ is an oblique projection of the plane. This also explains the 
Freeman relation, $L \propto R^2$ (\cite{fre70}), and the 
radius--velocity relation (\cite{tf77}). 
 
We used the same $I$-band data-set to fit equation~(3). In practice, we 
define $ k = \beta/\alpha$ and fit equation~(3) by varying $\alpha$ and 
$k$. Obviously, equation~(3) becomes the original Tully--Fisher relation 
when $ k = 0.0$, or it goes to the scaling plane of Koda et al. (2000a)
when $k = 1.0$.  After excluding those galaxies (1) whose recession velocities 
deviate by more than 1000 km~s$^{-1}$ from the mean velocity of a 
cluster, (2) without \HI measurements and (3) which may not be a 
cluster member, we finally obtained 160 galaxies in the sample. Throughout 
this paper $H_0 =$ 75 km~s$^{-1}$Mpc$^{-1}$ is assumed. We first used the 
$I$-band {\it total} absolute magnitude data, $M=M_I^{\rm tot}$, together 
with $V = W_{20}$ and $R = R_{I23.5}$ (in kpc), and obtained the minimum 
$\chi^2=252.2$ at $(\alpha, \beta, \gamma)$ = $(-4.05\pm0.21$, 
$-3.00\pm0.14$, $-8.01\pm0.43)$, see figure~\ref{han92chi}. This minimum 
$\chi^2$ value is much smaller than the minimum $\chi^2=697.6$
from the Tully--Fisher relation ($k=0$). It can also be seen in 
figure~\ref{han92chi} that the data scatter is much smaller when the 
galactic radius is involved. The best-fitting value of $k= \beta/\alpha$ is 
$0.74\pm0.06$, i.e. it is at neither $k=0.0$ nor $k=1.0$ 
(figure~\ref{han92chi}).  All fitting results, including those found 
below, are listed in table~\ref{bigtab}. The last two columns are the
$\chi^2$ values and the scattering of luminosity $\delta$, both for the
original Tully--Fisher relation and for the best fitting plane.
 
To account for light within the linear radius, the absolute magnitude, 
$M = M_I^{23.5}$, should probably be used.  We then obtained 
the best-fitting parameters: $ (\alpha, \beta, \gamma)$ = 
($-4.75\pm0.21$, $-2.85\pm0.14$, $-6.21\pm0.43$) at $k = 0.60\pm0.04$ 
and $\chi^2 = 359.4$.
 
\begin{table*} 
\begin{small} 
\caption{Search for the fundamental plane from observational data.
\label{bigtab}} 
\setlength{\tabcolsep}{1.5mm}
\begin{tabular}{llllccccccc} 
\hline 
\hline 
Sources & \multicolumn{3}{c}{Input data} & \multicolumn{1}{c}{Number} 
&\multicolumn{5}{c}{Fitting results}  &$\delta_{\rm TF}/\delta$\\ 
  \cline{2-4} \cline{6-10} 
  &  $L$  &  $V$ &  $R$  & of objects &  $\alpha$  & $\beta$ &  $\gamma$ &  
$k$  & $\chi^2_{\rm TF} / \chi^2$& (mag) \\ 
\hline 
Han92-I-1&$M_I^{\rm tot}$&$W_{20}$&$R_{I23.5}$&160&$-4.05\pm0.21$&$-3.00\pm0.14$&$-8.01\pm0.43$&$0.74\pm0.06$&697.6/252.2&0.42/0.25\\ 
Han92-I-2&$M_I^{23.5}$&$W_{20}$&$R_{I23.5}$&160&$-4.75\pm0.21$&$-2.85\pm0.14$&$-6.21\pm0.43$&$0.60\pm0.04$&760.0/359.4&0.44/0.30\\[1mm] 
HST-I-1&$M_I^{\rm tot}$&$W_{20}$&$R_{I23.5}$&17&$-5.67\pm0.38$&$-2.88\pm0.29$&$-3.92\pm0.80$&$0.51\pm0.06$&124.8/~~17.8&0.32/0.15\\ 
HST-I-2&$M_I^{\rm tot}$&$W_{20}$&$R_{I25}$&17&$-6.84\pm0.34$&$-2.04\pm0.25$&$-1.63\pm0.74$&$0.30\pm0.04$&124.8/~~56.8&0.32/0.24\\ 
HST-R-1&$M_R^{\rm tot}$&$W_{20}$&$R_{R23}$&17&$-5.30\pm0.38$&$-2.26\pm0.25$&$-5.37\pm0.81$&$0.43\pm0.06$&113.7/~~29.2&0.32/0.17\\ 
HST-R-2&$M_R^{\rm tot}$&$W_{20}$&$R_{R25}$&17&$-6.19\pm0.32$&$-2.10\pm0.25$&$-2.75\pm0.68$&$0.34\pm0.04$&113.7/~~42.2&0.32/0.21\\ 
HST-V-1&$M_V^{\rm tot}$&$W_{20}$&$R_{V23}$&17&$-4.16\pm0.41$&$-2.54\pm0.23$&$-7.86\pm0.91$&$0.61\pm0.08$&144.7/~~23.0&0.35/0.16\\ 
HST-V-2&$M_V^{\rm tot}$&$W_{20}$&$R_{V25}$&17&$-5.36\pm0.34$&$-2.59\pm0.25$&$-4.04\pm0.74$&$0.48\pm0.06$&144.7/~~32.9&0.48/0.19\\ 
HST-B-1&$M_B^{\rm tot}$&$W_{20}$&$R_{B23}$&17&$-3.85\pm0.44$&$-2.30\pm0.26$&$-8.62\pm0.99$&$0.60\pm0.10$&124.5/~~47.0&0.41/0.25\\ 
HST-B-2&$M_B^{\rm tot}$&$W_{20}$&$R_{B25}$&17&$-4.19\pm0.39$&$-3.41\pm0.33$&$-5.69\pm0.85$&$0.81\pm0.10$&124.5/~~18.9&0.41/0.18\\[1mm] 
PW00-I-1&$M_I^{\rm tot}$&$2\langle V\rangle$&$R_{I23.5}$&74&$-2.56\pm0.26$&$-3.96\pm0.24$&$-11.1\pm0.51$&$1.55\pm0.18$&371.8/~~88.5&0.51/0.22\\ 
PW00-I-2&$M_I^{\rm tot}$&$2\langle V\rangle$&$r_{\rm d}$&74&$-5.16\pm0.19$&$-1.79\pm0.16$&$-7.88\pm0.45$&$0.35\pm0.03$&371.8/252.9&0.51/0.38\\[1mm] 
C96-R-1&$M_R^{\rm tot}$&$V_{2.2}$&2.15$r_{\rm d}$&285&$-4.11\pm0.19$&$-2.23\pm0.12$&$-8.45\pm0.41$&$0.54\pm0,04$&703.5/359.4&0.47/0.34\\ 
C96-R-2&$M_R^{\rm tot}$&$V_{2.2}$&$R_{R23}$&285&$-2.25\pm0.23$&$-3.64\pm0.16$&$-11.4\pm0.46$&$1.62\pm0.18$&703.5/183.3&0.47/0.24\\ 
\hline 
\hline 
\end{tabular} 
\end{small} 
\end{table*}

 
\begin{figure*}
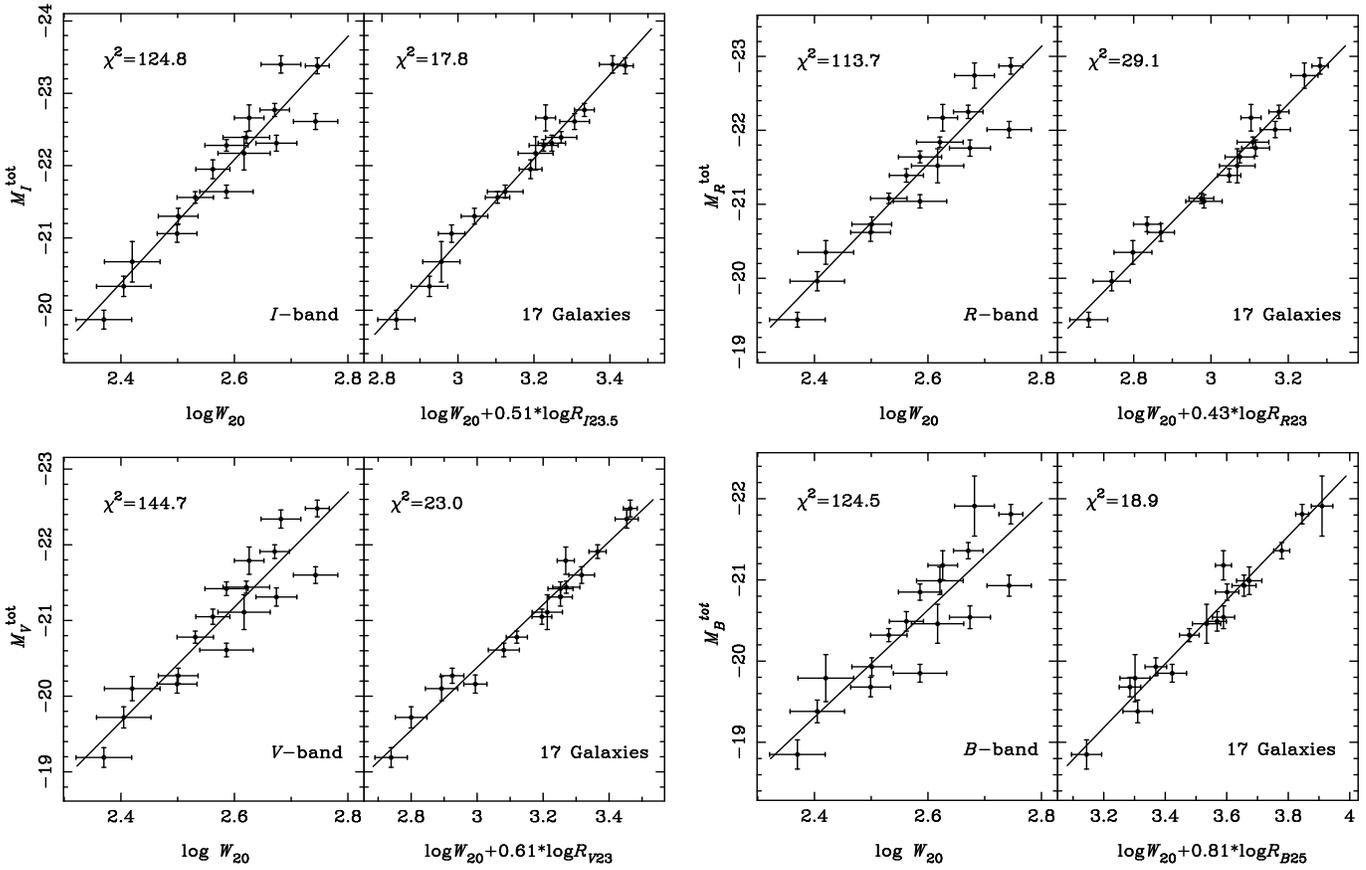
         
\centerline{\begin{tabular}{cc}
\mbox{\psfig{file=JLHan.fig2a.ps,width=88mm,height=56.5mm,angle=270}} & 
\mbox{\psfig{file=JLHan.fig2b.ps,width=88mm,height=55mm,angle=270}} \\[2mm] 
\mbox{\psfig{file=JLHan.fig2c.ps,width=88mm,height=55mm,angle=270}} & 
\mbox{\psfig{file=JLHan.fig2d.ps,width=88mm,height=55mm,angle=270}} \\ 
\end{tabular}} 
\caption{Scattering from the Tully--Fisher relation (left) 
and the best-fitting plane (right) for the Hubble calibration 
galaxies at the $IRVB$ bands. Data were taken from Sakai et al. (2000) and 
Macri et al. (2000). 
\label{hb00chi}} 
\end{figure*} 
 
%

\section{The Fundamental Plane in Other Bands} 
As a matter of fact, not many publications about the surface photometry 
of spiral galaxies contain information about galactic sizes. The database 
of Tully--Fisher calibration galaxies (\cite{mhs+00}; \cite{smh+00}), 
the database of Courteau (1996, 1997)\nocite{cou96}\nocite{cou97}, and 
the very recently published $I$-band data by Palunas and Williams (2000) 
\nocite{pw00} were used to search for the fundamental plane.
 
\subsection{The Hubble Calibration Galaxies} 
Sakai et al. (2000)\nocite{smh+00} published the Cepheid distances and 
carefully-corrected absolute magnitudes at $BVRIH$-bands of 21 calibration 
galaxies for the Hubble Space Telescope (HST) key project for the distance 
scale. They also presented the well measured and corrected \HI 20\% line 
widths. Macri et al. (2000)\nocite{mhs+00} further published the brightness 
profiles at the $BVRI$-bands and isophotal radii $R_{B25}$ and 
$R_{I23.5}$. We measured the the isophotal radii ($R_{R23}$, 
$R_{R25}$, $R_{V23}$, $R_{V25}$, $R_{B23}$, and $R_{I25}$) from
the brightness profiles. Finally, there are 17 galaxies with all
parameters which we needed.
 
We fitted equation (3) to these observational parameters in each band. 
As expected, involving the radius can greatly reduce the scattering 
around the Tully--Fisher relation (figure~\ref{hb00chi}). The $k$ values 
are around 0.5 (see lines with HST- in table~\ref{bigtab}). Notice that
in these HST data-sets, 
those with $R_{I23.5}$ and $R_{B25}$ produced the smallest residuals. 
 
\subsection{$I$-Band Data of Palunas and Williams (2000)} 
  
Very recently, Palunas and Williams (2000)\nocite{pw00} conducted $I$-band 
photometry and H$\alpha$ measurements of 74 galaxies. They presented 
the corrected magnitudes, $M_I^{\rm tot}$ and radii, $R_{I23.5}$, at the 
23.5~mag~arcsec$^{-2}$ isophote. They showed that for Freeman type--I
galaxies, i.e., galaxies with an exponential disk, the disk scale length,
$r_{\rm d}$, is very well--correlated with the disk size of $R_{I23.5}$,
while this is not the case for Freeman type--II galaxies with a large
fraction of disk which does not follow the exponential disk distribution
(see their figure~2). For the Tully--Fisher relation, they measured 
the rotation speed from the weighted average of the rotation curve points, 
and defined the velocity width as twice the speed. We corrected 
these widths for the inclination angle and redshift effect. 
 
\begin{figure*}    
\centerline{\begin{tabular}{rl} 
\mbox{\psfig{file=JLHan.fig3a.ps,width=110mm,height=60mm,angle=270}} & 
\mbox{\psfig{file=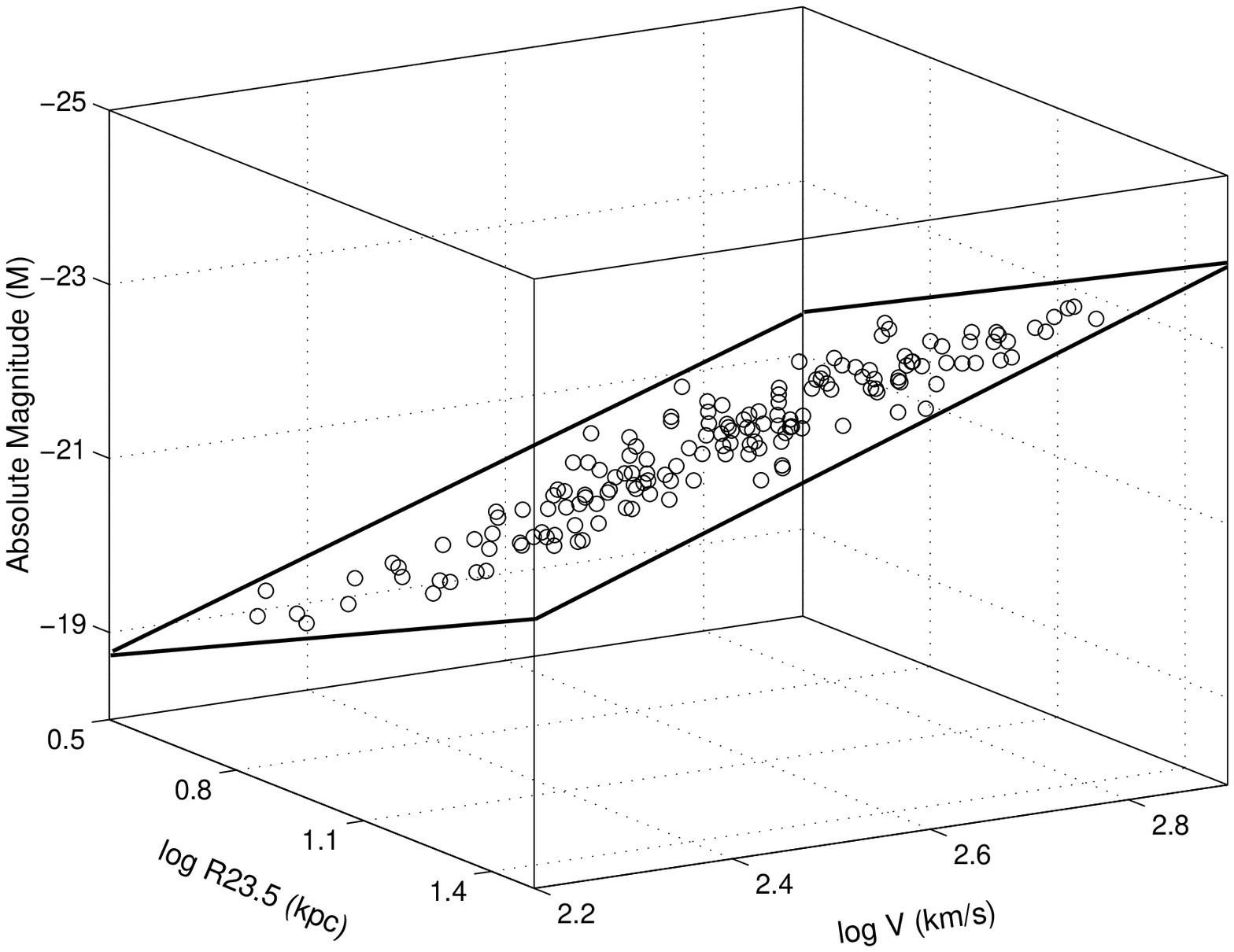,width=70mm,height=60mm,angle=0}} 
\end{tabular}}
\caption{Comparison of the Tully--Fisher relation (the left side) and 
the best-fitting plane (the right side). Data of 74 galaxies were taken from 
Palunas and Williams (2000). The typical rms errors of the absolute magnitudes 
were taken as 0.2 mag. The data distribution in 3-D space was also plotted.
\label{pw00chi} 
} 
\end{figure*}

As can be seen in figure~\ref{pw00chi}, we found that the fitting with 
$R_{I23.5}$ can largely diminish the scattering around the Tully--Fisher
relation.  See table~\ref{bigtab} (see lines with PW00) for the fitting
results. When the disk-scale length, $r_{\rm d}$, was used, the
 residual could be maximally reduced when $k=0.35$.
 
\begin{figure*}
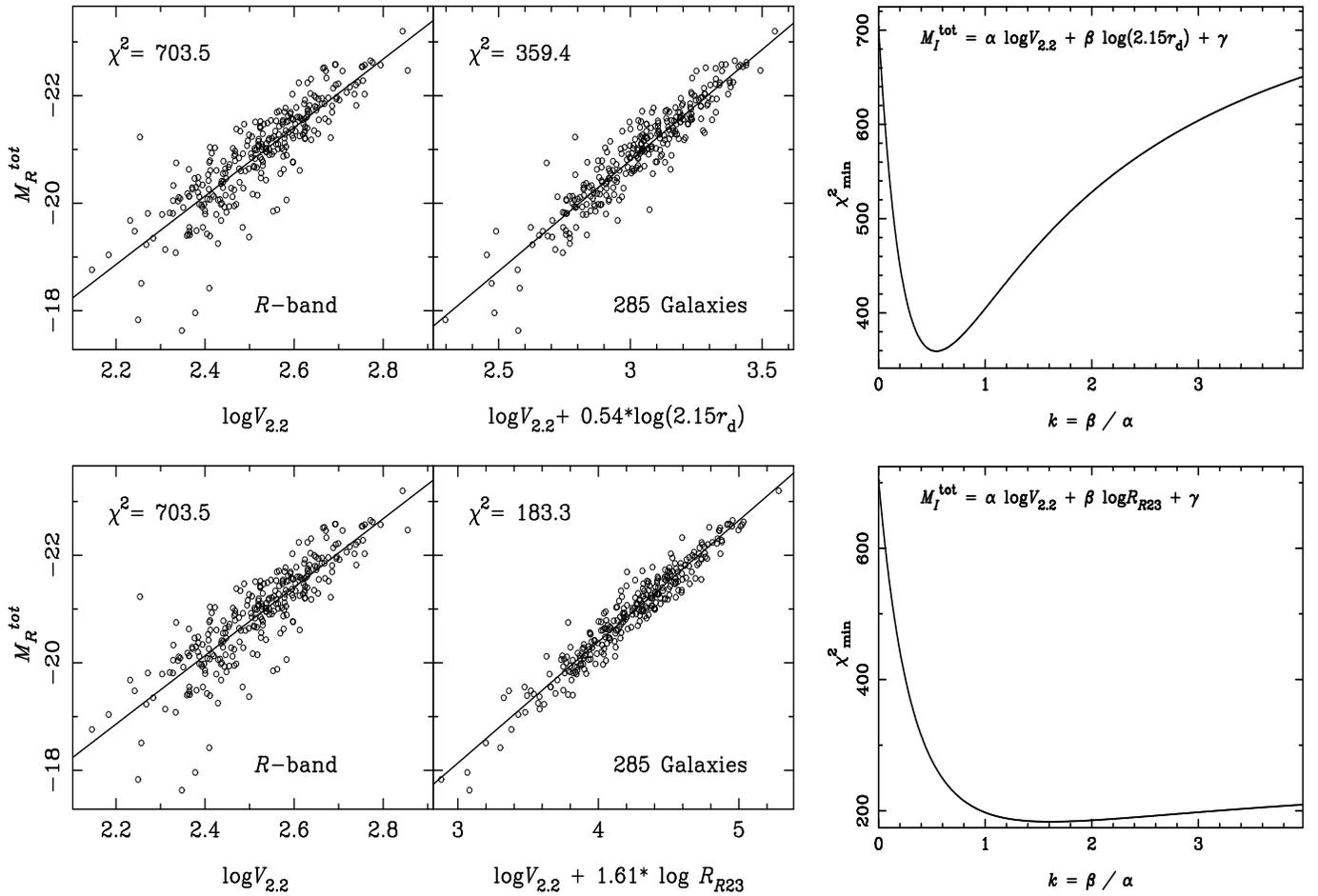
        
\centerline{\begin{tabular}{cc} 
\mbox{\psfig{file=JLHan.fig4a.ps,width=110mm,height=60mm,angle=270}} & 
\mbox{\psfig{file=JLHan.fig4b.ps,width=67mm,height=60mm,angle=270}} \\[3mm] 
\mbox{\psfig{file=JLHan.fig4c.ps,width=110mm,height=60mm,angle=270}} & 
\mbox{\psfig{file=JLHan.fig4d.ps,width=67mm,height=60mm,angle=270}} 
\end{tabular}}
\caption{Comparison of the Tully--Fisher relation (the left side) and 
the best plane (the right side). Data of 285 galaxies were taken from 
Courteau (1996,1997). The typical rms errors of the absolute magnitudes 
were taken as 0.3 mag. The variation in the fitting residual $\chi^2$  
with the ratio $k=\beta/\alpha$ is plotted on the right side. 
\label{cou97}
} 
\end{figure*} 
\subsection{$R$-Band Data of Courteau (1996, 1997)} 

\setcounter{footnote}{6} 

Courteau (1996, 1997)\nocite{cou96}\nocite{cou97} published the 
kinematic (from optical spectroscopic observations) and photomatic 
(at $R$-band) data of 304 late-type spiral galaxies. 
Besides the isophotal 
radii, $R_{R23}$ and $R_{R25}$, he also obtained a scale length for 
exponential disks, $r_{\rm d}$, to express the disk size. Courteau
(\yearcite{cou97}) found that the peak rotational velocity, $V_{2.2}$,
measured at $R = 2.15r_{\rm d}$ matches the 21 cm \HI line-widths 
best, and yields the smallest Tully--Fisher residuals. Therefore, we
use the velocity, $V_{2.2}$, for our discussions below. We took these
parameters\footnote{ 
Courteau (1996, 1997) used the Hubble constant of  $H_0 =$ 70 
km~s$^{-1}$Mpc$^{-1}$ to calculate the absolute size. This is fine 
for our discussion without further correction.},
together with the corrected total absolute magnitude,
$M_R^{\rm tot}$, from his combined data set (Courteau 1999, astro-ph/9903297). 

As listed in table~\ref{bigtab} (lines with C96-), using the
scale-length to express the disk size 
gives the minimum at $k\sim0.5$, but the residual is reduced only by 
a factor of 2 compared with the Tully--Fisher relation.  See 
figure~\ref{cou97} for illustrations.  In contrast, using the isophotal
sizes of galaxies can achieve a much tighter correlation with $k>1.0$.
We discuss this point later.
 
\section{Discussion and Conclusions} 

\subsection{The Plane is Fundamental}

We have confirmed the conclusion by Koda et al. (2000a) that the
galactic radius is a fundamental parameter which should be considered
to form the best--fitting plane in three-parameter space. As one can
see from table 1, except for the cases in which both the optically
measured velocities ($V_{2.2}$  or $\langle V\rangle$) and the isophotal
radii are used together, the $\beta$ values are always around $-2.5$,
while the values of $\alpha$ are always around $-5.0$. Thus,
approximately, $M = -5.0 \log V -2.5 \log R + \gamma$. Since
$M=-2.5\log L$, we have $$L = c V^2R.$$ Here, $c$ is a constant related to
$\gamma$, which greatly varies for different color bands or different
expression of luminosities; $V^2R$ is expected from the virial theorem.
The total mass, including the contributions from both the
disk and the dark matter, should satisfy 
$ {\mathcal M }(\le r) = [V^2_{\rm rot}(r)\;r]/G$, so that
the rotational velocity, $V_{\rm rot}(r)$, at radius
$r$ is related to the total mass, ${\mathcal M}$, inside the radius. 
where, $G$ is the gravitational constant. Therefore, our results imply that
the total luminosity is proportional, on average, to the total mass
inside the galactic radius. We conclude from an observational
point of view that the plane represented  by $L\sim V^2R$
is fundamental for spiral galaxies.

The mass of spiral galaxies is mostly in the form of
dark matter, probably in the halo, while the luminosity is given
mostly by the number of stars in the galaxy disk (see figure 5).
Therefore, the fundamental
plane derived by us implies a tighter relation between the bright mass
in the disk and the dark matter in the halo; at least they have
similar distribution profiles. This is coincident
with the conclusion derived from the universal rotation curve 
by \citet{sfp93} and \citet{pss96} on the tight coupling between
the dark and the luminous matter in spiral galaxies.
In general, the luminosity
can be expressed as $L= (m_{\rm d}/\Upsilon ){\mathcal M}$, where
$m_{\rm d}$ is the fraction
of disk mass in the total mass, and $\Upsilon$ is the mass-to-light ratio.
Our result seems to favor the constant $m_{\rm d}$ and $\Upsilon$ values for 
spiral galaxies, which are often assumed in theoretical studies
(see Mo et al. 1998; Shen et al. 2001).\nocite{mmw98}\nocite{sms01}
In addition, we noted that the best-fitting planes (the $k$ values) 
for spiral galaxies in our selected samples vary only slightly in
different wavebands, which may result from the formation history of the 
galaxies, and can be used to constrain theories of galaxy formation. 
 
\subsection{Tully--Fisher Residuals and Galactic Radius}

Note that the isophotal radii are more effective than the disk
scale-length in reducing the scatter. As can be seen from table 1,
the photometry-limited observational parameters, such as 
$R_{I23.5}$ from Han (1992)\nocite{han92} and Palunas and Williams 
(2000)\nocite{pw00}, $R_{R23}$ from Courteau (1996)\nocite{cou96}, 
and $R_{B25}$ for the Hubble calibration galaxies, can work better.
However, when the optically measured velocity is used together with
the isophotal radii, the best-fitting
planes are at $k>1$. Two independent data--sets of Palunas and Williams
(2000) and Courteau (1996,1997) show qualitatively the same result.
Although we noticed the fact that the two observational parameters,
$\log(V_{2.2})$ and $\log(R_{23.5})$, are measured at distinctly different
radii, this may not be the reason for the flat variation of $\chi^2$
when $k>1$. 

The fundamental plane involving the linear galactic size can reduce
the Tully--Fisher residuals by about 50\%. The unified plane
$L \propto (VR)^{1.3}$ that Koda et al. (2000a)\nocite{ksw00a} found
from the $I$-band data-set can also work much  better than the Tully--Fisher
relation, though it may not be the fundamental plane.
Our results (in our table 1) are consistent with case $c$ of
theoretical simulations of Shen et al. (2001)\nocite{sms01},
in which they obtained the smallest scattering (see their table 1).

\subsection{Other Issues}

The fundamental plane does not help to improve the
distance estimates of galaxies using the luminosity, because the galactic
size here is a quantity having a distance dependence. In fact, during the above
searches for the plane, the effect as well as uncertainty of the galactic
distance was diminished for the fundamental plane, compared with that
in the Tully--Fisher relation.

Based on the Tully--Fisher relation ($L \propto V^4$) and 
Freeman's law ($R \propto L^{0.5}$), the fundamental plane described
by equation (3) requires $\alpha$ and $\beta$ to satisfy the relation
 $\alpha +2 \beta =-10$. If $\alpha = \beta$ (see discussion
by \cite{zh00}), one may find that $\alpha = \beta = -10/3$
and exactly naturally becomes $L \propto (VR)^{4/3}$, identical
to that obtained
by Koda et al. (2000a)\nocite{ksw00a}.  Whenever $\alpha \not= \beta $,
the possible values of $\alpha$ and $\beta$ are in the ranges 
(0, $-$10) and (0, $-$5), respectively. The relation of \citet{wil99}
obtained from the $R$-band,
$v_{\rm TF}\propto L^{0.28}I_{\rm e}^{0.14}$, if converted to our manner,
is $L\simeq V^{7/3}R^{2/3}$. As can be seen from the search
results in table 1, these values slightly depend on the color band
and the sample selection. It is interesting
to note that these best-fitting $\alpha$ and $\beta$ values are 
always around the central values in their possible ranges. 
Both of the planes derived by Koda et al. (2000a)
and by us can recover the Tully--Fisher relation and the Freeman's law,
which are just projections onto 2-D of the fundamental plane in 3-D.

\begin{figure}
\psfig{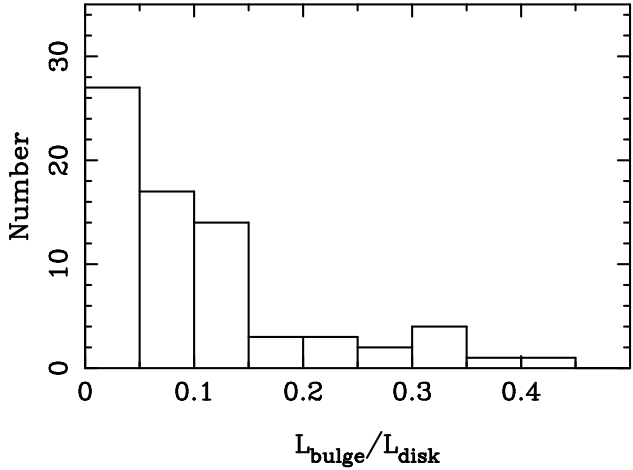}
\caption{Bulge to the disk luminosity ratio. Most late-type spiral galaxies 
do not have a very bright bulge. (Data from \cite{pw00}). 
\label{b2d}}
\end{figure}

In the above analyses, the bulge and disk parts of the spirals were
not separately considered. However, for late--type spiral galaxies, 
the bulges often have rather smaller luminosities than the disks 
(figure~\ref{b2d}). In our analysis, most galaxies are of the late--type,
and thus the effect of the bulge component on our results is fairly
small. The disk contribution to the luminosity is dominant.
  
\subsection{Conclusions}
Our search results reveal that there exists a fundamental plane for
spiral galaxies, which can be expressed by $L\sim V^2R$.
The galactic physical size should be involved to form the fundamental
plane in three--dimensional space of $\log L$ , $\log R$, and $\log V$. 
The size should be preferably the linear radius at a given isophotal
limit if the velocity is taken from the width of H~{\sc i} gas. Otherwise
the scale length of optical disk can be used as $R$ if the rotation velocity
is measured from the optical spectrum. This fundamental plane can
reduce the residual of the Tully--Fisher relation by an amount of
about 50\%, implying that only the other 50\% is attributed by the
measurement uncertainties. The fundamental plane exists in all 
optical bands. Such a fundamental plane is probably related to the mass and
mass distribution of spiral galaxies, and should be used to test theoretical
work concerning galaxy formation.

\bigskip
This work is supported partially by the National Natural Science 
Foundation of China, the National Key Basic Research Science
Foundation (NKBRSF G19990752), and the One-Hundred-Talent Program.


\begin{thebibliography}{}

\bibitem[{Avila-Reese}, {Firmani} (2000)]{af00}
{Avila-Reese}, V., \& {Firmani}, C. 2000, { Rev. Mex. Astronom. Astrofis.}, {\rm 36}, 23

\bibitem[{Broeils}, {Rhee} (1997)]{br97}
{Broeils}, A.~H., \& {Rhee}, M.-H. 1997, { \aap}, {\rm 324}, 877

\bibitem[{Courteau} (1996)]{cou96}
{Courteau}, S. 1996, { \apjs}, {\rm 103}, 363

\bibitem[{Courteau} (1997)]{cou97}
{Courteau}, S. 1997, { \aj}, {\rm 114}, 2402

\bibitem[{Courteau} and {Rix} (1999)]{cr99}
{Courteau}, S., \& {Rix}, H. W. 1999, { \apj}, {\rm 513}, 561

\bibitem[{Dalcanton} et al. (1997)]{dss97}
{Dalcanton}, J.~J., {Spergel}, D.~N., \& {Summers}, F.~J. 1997, { \apj}, {\rm
  482}, 659

\bibitem[{Djorgovski}, {Davis} (1987)]{dd87}
{Djorgovski}, S., \& {Davis}, M. 1987, { \apj}, {\rm 313}, 59

\bibitem[{Dressler} \etal  (1987)]{dlb+87}
{Dressler}, A., {Lynden-Bell}, D., {Burstein}, D., {Davies}, R.~L., {Faber},
  S.~M., {Terlevich}, R., \& {Wegner}, G. 1987, { \apj}, {\rm 313}, 42

\bibitem[{Eisenstein}, {Loeb} (1996)]{el96}
{Eisenstein}, D.~J., \& {Loeb}, A. 1996, { \apj}, {\rm 459}, 432

\bibitem[{Firmani}, {Avila-Reese} (2000)]{fa00}
{Firmani}, C., \& {Avila-Reese}, V. 2000, { \mnras}, {\rm 315}, 457

\bibitem[{Freeman} (1970)]{fre70}
{Freeman}, K.~C. 1970, { \apj}, {\rm 160}, 811

\bibitem[{Giovanelli} et al. (1997)]{ghh+97}
{Giovanelli}, R., {Haynes}, M.~P., {Herter}, T., {Vogt}, N.~P., 
        {da Costa}, L.~N., {Freudling}, W., {Salzer}, J.~J., \& 
        {Wegner}, G. 1997, { \aj}, {\rm 113}, 53

\bibitem[{Han} (1992)]{han92}
{Han}, M. 1992, { \apjs}, {\rm 81}, 35

\bibitem[{Koda}, {Sofue}, \& {Wada} (2000a)]{ksw00a}
{Koda}, J., {Sofue}, Y., \& {Wada}, K. 2000a, { \apj}, {\rm 531}, L17

\bibitem[{Koda}, {Sofue}, \& {Wada} (2000b)]{ksw00b}
{Koda}, J., {Sofue}, Y., \& {Wada}, K. 2000b, { \apj}, {\rm 532}, 214

\bibitem[{Kodaira} (1989)]{kod89}
{Kodaira}, K. 1989, { \apj}, {\rm 342}, 122

\bibitem[{Macri} \etal  (2000)]{mhs+00}
{Macri}, L.~M., {Huchra}, J.~P., {Sakai}, S., {Mould}, J.~R., \& {Hughes}, S.
  M.~G. 2000, { \apjs}, {\rm 128}, 461

\bibitem[{Mo}, {Mao} (2000)]{mm00}
{Mo}, H.~J., \& {Mao}, S. 2000, { \mnras}, {\rm 318}, 163

\bibitem[{Mo} et al. (1998)]{mmw98}
{Mo}, H.~J., {Mao}, S., \& {White}, S. D.~M. 1998, { \mnras}, {\rm 295}, 319

\bibitem[{Palunas}, {Williams} (2000)]{pw00}
{Palunas}, P., \& {Williams}, T.~B. 2000, { \aj}, {\rm 120}, 2884

\bibitem[{Persic} \etal (1996)]{pss96}
{Persic}, M., {Salucci}, P., \& {Stel}, F. 1996, { \mnras}, {\rm 281}, 27

\bibitem[Sakai \etal  (2000)]{smh+00}
Sakai, S., Mould, J.R., Hughes, S.M.G., Huchra, J.R., Macri, L.M., 
Kennicutt, R.C., Jr., Gibson, B.K., Ferrarese, L., et al. 2000, { \apj}, 
{\rm 529}, 698

\bibitem[{Salucci} \etal (1993)]{sfp93}
{Salucci}, P., {Frenk}, C. S., \& {Persic}, M. 1993, { \mnras}, {\rm 262}, 392

\bibitem[{Shen} \etal (2001)]{sms01}
{Shen}, S., {Mo}, H. J., \& {Shu}, C. 2001, { \mnras}, {\rm submitted}, astro/ph-0105095

\bibitem[{Silk} (1997)]{sil97}
{Silk}, J. 1997, { \apj}, {\rm 481}, 703

\bibitem[{Steinmetz}, {Navarro} (1999)]{sn99}
{Steinmetz}, M., \& {Navarro}, J.~F. 1999, { \apj}, {\rm 513}, 555

\bibitem[{Tully}, {Fisher} (1977)]{tf77}
{Tully}, R.~B., \& {Fisher}, J.~R. 1977, { \aap}, {\rm 54}, 661

\bibitem[{Tully}, {Pierce} (2000)]{tp00}
{Tully}, R.~B., \& {Pierce}, M.~J. 2000, { \apj}, {\rm 533}, 744

\bibitem[{van den Bosch} (2000)]{vdb00}
{van den Bosch}, F.~C.\ 2000, { \apj}, {\rm 530}, 177

\bibitem[{Willick} (1999)]{wil99}
{Willick}, J.~A.\ 1999, { \apj}, {\rm 516}, 47

\bibitem[{Willick} \etal  (1997)]{wcf+97}
{Willick}, J.~A., {Courteau}, S., {Faber}, S.~M., {Burstein}, D., {Dekel}, A.,
  \& {Strauss}, M.~A.\ 1997, { \apjs}, {\rm 109}, 333

\bibitem[{Zou}, {Han} (2000)]{zh00}
{Zou}, Z.~L., \& {Han}, J.~L. 2000, { Chinese Phys. Lett.}, {\rm 17}, 935

\end{thebibliography}
\end{document}